\documentclass[twocolumn,aps,prl,showpacs,superscriptaddress]{revtex4}
\usepackage{graphicx}
\usepackage{amsmath}
\usepackage{amsfonts}
\usepackage{amssymb}

\begin{document}

\author{Arne Brataas}
\affiliation{Department of Physics, Norwegian University of Science and Technology,
N-7491 Trondheim, Norway}
\author{Yaroslav Tserkovnyak}
\affiliation{Lyman Laboratory of Physics, Harvard University, Cambridge, Massachusetts
02138, USA}
\title{Spin and Charge Pumping by Ferromagnetic-Superconductor Order
Parameters}

\begin{abstract}
We study transport in ferromagnetic-superconductor/normal-metal
systems. It is shown that charge and spin currents are pumped from
ferromagnetic superconductors into adjacent normal metals by adiabatic changes in the order parameters induced by external electromagnetic
fields. Spin and charge pumping identify the symmetry of the superconducting
order parameter, e.g., singlet pairing or triplet pairing with opposite or
equal spin pairing. Consequences for ferromagnetic-resonance
experiments are discussed.
\end{abstract}

\date{\today}
\maketitle


Ferromagnetism induces a spin-dependent asymmetry in the densities of
itinerant carriers. In contrast, superconductivity pairs electrons with
equal or opposite spins depending on the symmetry of its order parameter.
The coexistence of the two order parameters has been considered
to be a rare phenomenon. However, recent experimental progress has
demonstrated that ferromagnetism and superconductivity coexist in some
materials like RuSr$_{2}$GdCu$_{2}$O$_{8}$ \cite{Tallon:ieee99}, UGe$_{2}$ 
\cite{Saxena:nat00}, ZrZn$_{2}$ \cite{Pfleiderer:nat01}, and URhGe \cite%
{Aoki:nat01}. The experiments find triplet pairing in URhGe, and strong
indications of triplet pairing in UGe$_{2}$ and ZrZn$_{2}$; they
furthermore suggest that the same electrons are responsible for
ferromagnetism as well as superconductivity. Besides, ferromagnetism and superconductivity are predicted to be simultaneously induced
in hybrid ferromagnet (F)/normal-metal (N)/superconductor (S)
systems \cite{Huertas:prl02}. These recent experimental discoveries, and the
possibility of tailoring superconductivity and ferromagnetism in nanoscale
systems, enable exploring novel physics involving pairing and spin-related
transport processes. A variety of interesting spin phenomena have already
been observed in hybrid F/N and semiconductor systems: e.g.,
giant-magnetoresistance|related effects, spin precession,
and current-induced magnetization dynamics \cite{GMROptics}.
It is therefore natural to expect that interesting rich phenomena should
also occur in FS/N systems.

This Letter demonstrates how the coexistence of superconductivity and
ferromagnetism is manifested in the adiabatic pumping in hybrid FS/N
structures. In particular, we study spin and charge
pumping when the magnetization slowly precesses, which can be achieved in ferromagnetic resonance (FMR) and in current-induced magnetization dynamics. FMR
experiments have already been carried out to investigate the magnetism in
RuSr$_{2}$GdCu$_{2}$O$_{8}$ \cite{Fainstein:prb99}. We also consider pumping
by slow variations in the phase of the singlet or triplet order parameter, or
in the direction of the triplet order parameter. These can be induced by
electric and magnetic fields to be discussed below. By pumping we thus mean
the spin and charge flows into the adjacent normal conductors in response to
adiabatic changes in the FS order parameters. Consequently, in the case of
pumping by the magnetization direction, we compute the spin current $\mathbf{%
I}_{s}$ and the charge current $I_{c}$ for a given rate of the
magnetization-direction change, and for pumping by changing pairing, we
compute the same quantities as functions of the phase-change or the
direction-change rates of the pair correlations.
We employ two approaches giving identical results:
1) solving the time-dependent ac problem directly and 2) using a gauge
transformation to obtain a time-independent dc problem. We first explain
our model and notation, before proceeding to the derivation and results.
Experimental consequences are discussed in the end.

A ferromagnetic superconductor is treated in the mean-field approximation,
where ferromagnetism is represented by the average magnetization and superconductivity is described by a pair
potential. Our model is phenomenological and we do not discuss the
microscopic origin of the exchange field or the superconducting pairing. The
Bogoliubov-de Gennes (BdG) equation is
\begin{equation}
\left( 
\begin{array}{cc}
\hat{\xi} & \hat{\Delta} \\ 
-\hat{\Delta}^{\ast } & -\hat{\xi}^{\ast }%
\end{array}%
\right) \left( 
\begin{array}{c}
\hat{u} \\ 
\hat{v}%
\end{array}%
\right) =i\hbar \frac{\partial }{\partial t}\left( 
\begin{array}{c}
\hat{u} \\ 
\hat{v}%
\end{array}%
\right) \,,  \label{BdG}
\end{equation}%
where $\hat{\xi}=H_{0}\hat{1}+\boldsymbol{\hat{\sigma}}\cdot \boldsymbol{%
\epsilon }_{\text{xc}}$ is the single-particle Hamiltonian, $\boldsymbol{%
\epsilon }_{\text{xc}}=\epsilon _{\text{xc}}\mathbf{m}$ (with $\epsilon _{%
\text{xc}}>0$) is the ferromagnetic exchange field along the magnetization
direction $\mathbf{m}$ and $\hat{\Delta}=\left(d_{0}\hat{1}+%
\boldsymbol{\hat{\sigma}}\cdot \boldsymbol{d}\right) i\hat{\sigma}_{y}$
is the superconducting pair potential, in terms of the singlet (scalar)
part, $d_{0}$, and the triplet (vector) part, $\boldsymbol{d}%
=(d_{x},d_{y},d_{z})$. $\hat{u}^{T}=(u^{\uparrow
},u^{\downarrow })$ are spin-dependent electron wave functions and $\hat{v}%
^{T}=(v^{\uparrow },v^{\downarrow })$ are those of holes; $\boldsymbol{\hat{%
\sigma}}=\left( \hat{\sigma}_{x},\hat{\sigma}_{y},\hat{\sigma}_{z}\right) $
is the vector of Pauli matrices. The single-particle Hamiltonian $H_{0}$
contains the kinetic and potential-energy terms. The Fermi energy is taken
to be the largest relevant energy scale. The local exchange field can be
position dependent close to the interface, $\boldsymbol{\epsilon }_{\text{xc}%
}(\mathbf{r})$, and the pair potential $\hat{\Delta}(\mathbf{k},\mathbf{r})$
can also be position as well as wave-vector, $\mathbf{k}$, dependent \cite%
{Bruder:prb90}. For simplicity, we assume that the exchange field and the
pair potential are uniform inside the superconductor and drop to zero at the
FS/N interface: $\boldsymbol{\epsilon }_{\text{xc}}(\mathbf{r})=\boldsymbol{%
\epsilon }_{\text{xc}}\Theta (z)$ and $\hat{\Delta}(\mathbf{k},\mathbf{r})=%
\hat{\Delta}(\mathbf{k})\Theta (z)$, where $\Theta (z)$ is the Heaviside
step function and $z$ is the coordinate perpendicular to the FS/N interface.

Fermionic statistics dictates $\hat{\Delta}(\mathbf{k})=-\hat{\Delta}^{T}(-%
\mathbf{k})$ \cite{Sigrist:rmp91}. The singlet (triplet) part of $\hat{\Delta%
}$ thus needs to have even (odd) parity: $d_0(\mathbf{k})=d
_{0}(-\mathbf{k})$ and $\mathbf{d}(\mathbf{k})=-\mathbf{d}(-%
\mathbf{k})$. We study in the following two simple cases of triplet
superconductors: opposite-spin pairing (OSP) along the exchange field, $%
\mathbf{d}(\mathbf{k})\times \boldsymbol{\epsilon }_{\text{xc}}=0$ and
equal-spin pairing (ESP), $\mathbf{d}(\mathbf{k})\cdot \boldsymbol{%
\epsilon }_{\text{xc}}=0$ \cite{Powell:jpa03}. Triplet OSP superconductors
are described by a (complex-valued) vector $\boldsymbol{d}(\mathbf{k}%
)=d (\mathbf{k})\mathbf{m}$ along the magnetization direction. We show
that the transport properties in triplet OSP are similar to those of singlet
pairing. In ESP, superconductivity occurs independently for spins along and
opposite to the magnetization direction: By choosing the magnetization along
the $z$ axis, the superconducting pair potential decomposes into two terms,
corresponding to spins up and down along the $z$ axis,
$d^{\uparrow(\downarrow)}(\mathbf{k})=\mp d_{x}(\mathbf{k})+id_{y}(\mathbf{k})$.
Since superconducting correlations do not mix the spin-up and down
subsystems, two ESP phases can be distinguished: the A$_{1}^{\sigma }$
phase, where pairing occurs only for spin $\sigma $ (i.e., $d
^{\sigma }\neq 0$ and $d ^{-\sigma }=0$) and the A$_{2}$ phase, where
pairing occurs for both spins. A large exchange interaction, $\epsilon _{%
\text{xc}}\gtrsim |d_0|$ $(|d|)$, suppresses superconducting singlet (triplet)
OSP correlations~\cite{Powell:jpa03}. We therefore only consider $\epsilon _{%
\text{xc}}<|d_0|$ $(|d|)$ for these systems, so that the quasiparticle
excitations have a finite gap. Triplet ESP have a quasiparticle gap in the
superconducting spin channels independent of the size of the exchange
interaction \cite{Powell:jpa03}, and we do not make assumptions about the
ratio of the exchange field to the pair potential in this case.

\begin{figure}[tbp]
\includegraphics[width=\linewidth,clip=]{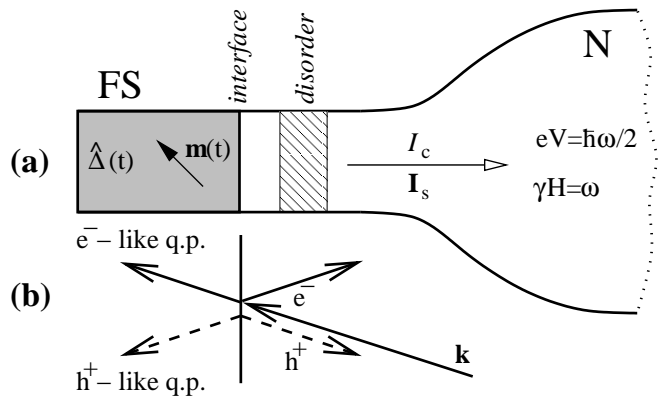}
\caption{(a) A ferromagnetic superconductor coupled
to a normal reservoir through a specular FS/N
interface in series with a normal disordered region. (b) Electrons incident
on the FS/N interface from the normal side
are reflected as electrons or holes and transmitted as
decaying electron- or hole-like quasiparticles.}
\label{sc}
\end{figure}

Let us apply the standard scattering-matrix approach \cite%
{Beenakker95,Brouwer:prb98,Brataas:prl00,Tserkovnyak:prl021,Blaauboer:prb02}
to an FS/N system. We assume that the Hamiltonian $H_{0}$ is continuous
across the FS/N interface and incorporate interfacial disorder and
band-structure mismatch into a \textquotedblleft
disordered\textquotedblright\ normal region \cite{Beenakker95}. Similarly to
Ref.~\cite{Beenakker95}, we solve the BdG equation for an electron (or hole)
incident on the specular FS/N interface from the normal-metal side. The
total reflection matrix is then found by concatenating the FS/N reflection
with the scattering by the normal disordered region for electrons and holes.
The FS layer, in series with the disordered region, is viewed as a scatterer
for electrons supplied by the normal reservoir, see Fig.~\ref{sc}. The
problem is simplified in the clean-superconductor limit, where the FS mean
free path is longer than the superconducting coherence length $\hbar
v_{F}/(\pi\Delta)$, expressed in terms of the Fermi velocity $%
v_{F}$ and quasiparticle gap $\Delta$. Focusing on the
low-temperature regime, $k_{B}T\ll\Delta$, we define the $%
M\times M$ spin-dependent electron$\rightarrow $electron and electron$%
\rightarrow $hole reflection matrices $r_{ee}^{\sigma }$ and $r_{he}^{\sigma
}$, where $M$ is the total number of quantum channels, or transverse
waveguide modes, at the Fermi level of the normal-metal lead and $\sigma $
is the spin label along the magnetization direction for incident electrons.
For singlet or triplet OSP, $r_{he}^{\sigma }$ describes reflection into
holes with spin $-\sigma $, while for triplet ESP, reflected holes have the
same spin $\sigma $ as incident electrons.

We find that for both dc and adiabatic ac transport, spin and charge
currents are uniquely determined by two real-valued conductances and one
complex-valued conductance, representing transport of spins aligned to the
magnetization, antialigned to the magnetization, and transverse to the
magnetization, respectively: 
\begin{eqnarray}
g^{\uparrow \uparrow } &=&\mbox{Tr}\left[ r_{he}^{\uparrow
}(r_{he}^{\uparrow })^{\dagger }\right] \,,\,\,g^{\downarrow \downarrow }=%
\mbox{Tr}\left[ r_{he}^{\downarrow }(r_{he}^{\downarrow })^{\dagger }\right]
\,,  \label{gA} \\
g^{\uparrow \downarrow } &=&\mbox{Tr}\left[ 1-r_{ee}^{\uparrow
}(r_{ee}^{\downarrow })^{\dagger }+(r_{he}^{\uparrow })^{\dagger
}r_{he}^{\downarrow }\right] \,.  \label{gmix}
\end{eqnarray}%
It is convenient to define the total conductance $g=g^{\uparrow \uparrow
}+g^{\downarrow \downarrow }$ and the polarization $p=(g^{\uparrow \uparrow
}-g^{\downarrow \downarrow })/g$. We will below interpret how these
conductances determine charge and spin flow, and discuss their values in various limits.

Let us first fix the magnetization direction $\mathbf{m}$ and consider
pumping by an adiabatically varying phase $\phi (t)$ of the superconducting
pair potential, $\partial _{t}\phi =\omega \ll (\epsilon _{\text{xc}%
},\Delta)/\hbar $. As is well known, this time-dependent problem can be
transformed into a dc problem by a gauge transformation
$U(t)=e^{i\phi (t)/2}$: Varying the phase
results in nonequilibrium transport corresponding to a dc response at a
voltage bias $\hbar\omega/(2e)$. The pumped current is thus nothing more
than a response to a voltage $V=\hbar\omega/(2e)$ applied between the
superconducting condensate and the normal metal, and we compute 
\begin{equation}
I_{c}=\frac{e}{2\pi}g\omega \,\,\,\mathrm{and}\,\,\,\mathbf{I}_{s}=-\frac{\hbar}{4\pi}pg\omega\mathbf{m}\,.  \label{Ic}
\end{equation}%
$g$ is thus the usual Andreev-reflection conductance, in units of
$e^{2}/(\pi\hbar)$. The spin current is determined by the conductance
polarization $p$. Eqs.~(\ref{Ic}) can also be derived in the ac pumping
framework of Ref.~\cite{Blaauboer:prb02}. 

Pure spin flow is generated by the variations in the
magnetization direction $\mathbf{m}$, induced by, e.g., a resonant rf
magnetic field. In Ref.~\cite{Tserkovnyak:prl021} we computed the
corresponding pumped spin flow into the normal metal, 
\begin{equation}
\mathbf{I}_{s}=\frac{\hbar}{4\pi}\left( \mbox{Re}g^{\uparrow \downarrow }\mathbf{m}\times\frac{d\mathbf{m}}{dt}+\mbox{Im}g^{\uparrow \downarrow }\frac{d\mathbf{m}}{dt}\right)\,,  \label{Is}
\end{equation}
for a general F/N contact with no superconducting correlations. Here
$g^{\uparrow\downarrow}=\mbox{Tr}[1-r_{ee}^{\uparrow}(r_{ee}^{\downarrow})^{\dagger}]$ is the F/N mixing conductance for transverse spins expressed in
terms of the normal-side reflection matrices \cite{Brataas:prl00}. The
pumped spin flow induces an enhanced Gilbert damping when the normal metal
is a good spin sink, so that $g^{\uparrow \downarrow }$ is experimentally
measurable \cite{Tserkovnyak:prl021}. We have generalized Eq.~(\ref{Is}) to
spin pumping by a superconducting ferromagnet: The pumped spin flow remains
of the form (\ref{Is}) with a redefined mixing conductance $g^{\uparrow
\downarrow }$ (\ref{gmix}). There is no accompanying charge pumping. We
derive Eqs.~(\ref{gmix}) and (\ref{Is}) by two different methods: First, we
extend the pumping approach of Ref.~\cite{Tserkovnyak:prl021}, where the
scattering matrix is time-dependent in spin space due to a slow variation in
the magnetization direction $\mathbf{m}$, to include electron$\rightarrow $%
hole reflection. Secondly, and much simpler, the calculation is reformulated
as a dc problem in the spin frame of reference which is moving together with 
$\mathbf{m}(t)$. The latter is achieved by (instantaneously) applying a
spin-rotation operator around the vector $\mathbf{m}\times \partial _{t}%
\mathbf{m}$ and correspondingly adding a new term in the Hamiltonian: $\hat{%
\xi}^{\prime }=-(\hbar /2)(\mathbf{m}\times \partial _{t}\mathbf{m})\cdot 
\boldsymbol{\hat{\sigma}}$. This term corresponds to an equilibrium
transverse spin accumulation which, in turn, induces the spin current (\ref%
{Is}). The mixing conductance (\ref{gmix}) is obtained after extending the
F/N dc theory of Ref.~\cite{Brataas:prl00} to account for electron$%
\rightarrow $hole reflection. By unitarity of the scattering matrix \cite%
{Beenakker95}, the real part of $g^{\uparrow \downarrow }$ is bounded from
above by twice the number of channels: $\mbox{Re}g^{\uparrow \downarrow
}\leq 2M$.

For triplet OSP superconductors, the derivation leading to Eq.~(\ref{Is})
assumes that the triplet pair-potential direction moves together with $%
\mathbf{m}$, while its wave-vector dependence is locked to the atomic
lattice: $\boldsymbol{d}(\mathbf{k},t)=d (\mathbf{k})\mathbf{m}(t)
$. For triplet ESP, the situation is more complex since the vector pair
potential $\boldsymbol{d}$ is perpendicular to $\mathbf{m}$, resulting
in an additional dynamic degree of freedom. In particular, the magnetization
motion does not uniquely determine the trajectory of $\boldsymbol{d}$.
In deriving Eq.~(\ref{Is}), we assumed that $\boldsymbol{d}$ rotates
together with $\mathbf{m}$ around $\mathbf{m}$'s instantaneous rotation
axis. We can get more complex trajectories by combining (instantaneous)
rotations of $\mathbf{m}$ and $\boldsymbol{d}$ with \textquotedblleft
twisting\textquotedblright\ of $\boldsymbol{d}$ around $\mathbf{m}$
and the overall phase variation of the pair potential. The induced
currents will then be given by adding the corresponding contributions to
pumping. Calculating the exact trajectories for realistic systems, which
might be governed by spin-orbit interactions in the lattice field or other
microscopic details, lies beyond the scope of this paper. A simple example
is the clockwise rotation of $\boldsymbol{d}$ around $\mathbf{m}$ with
frequency $\omega $ which induces additional currents 
\begin{equation}
I_{c}=-\frac{e}{2\pi}pg\omega \,\,\,\mathrm{and}\,\,\,\mathbf{I}_{s}=\frac{\hbar}{4\pi}g\omega\mathbf{m}\,.  \label{Ics}
\end{equation}%
Note that while the preceding equations are general, Eq.~(\ref{Ics})
applies to the case of triplet ESP only. For a FS/N system disconnected
from an Ohmic circuitry, the low-frequency charge flow vanishes, so that
the overall phase variation of $\boldsymbol{d}$ and its twisting
around $\mathbf{m}$ must result in canceling charge currents, $I_{c}$'s, in
Eqs.~(\ref{Ic}) and (\ref{Ics}), but a finite net spin current for $p<1$.
Eqs.~(\ref{Ics}) can be derived similarly to Eq.~(\ref{Is}),
either as a time-dependent pumping problem or a dc problem in the
gauge-transformed frame of reference rotating with $\boldsymbol{d}$.

Eqs.~(\ref{Ic}), (\ref{Is}), and (\ref{Ics}) are general expressions for
pumped charge and spin flows by varying FS order parameters. These currents
are all governed by two real-valued and one complex conductance parameters: $%
g^{\uparrow \uparrow }$, $g^{\downarrow \downarrow }$, and $g^{\uparrow
\downarrow }$, which can be evaluated in microscopic models. A finite value
of the electron$\rightarrow $hole reflection coefficient, $r_{he}^{\sigma }$%
, requires that an electron incident from the normal-metal reservoir gets
transmitted through the disordered region, converted into holes at the
interface, and transmitted back into the normal lead as a hole, see Fig.~\ref%
{sc}. For small (characteristic) transmission eigenvalues $T$
of the normal disordered region, $g^{\uparrow \uparrow }$ and
$g^{\downarrow \downarrow }$ therefore scale as $T^{2}$ rather than as $T$,
as in the Landauer-B\"{u}ttiker formula for an N/N contact.
Indeed, the Andreev conductance for
a singlet nonmagnetic-superconductor/normal-metal contact was shown to be $%
g_{\text{S/N}}=\sum_{m}2T_{m}^{2}/(2-T_{m})^2$, where $m$ labels the
transmission eigenvalues for scattering in the normal metal \cite%
{Beenakker95}. In the limit of no disorder (and no band-structure mismatch), 
$T_{m}\equiv 1$ and $g_{\text{S/N}}=2M$, i.e., Andreev reflection causes a
doubling of the conductance, as compared to the N/N interface. We generalize $g_{\text{S/N}}$ to the case of a magnetic singlet superconductor
with an $s$-wave symmetry of the pair potential, $d_0(\mathbf{k}%
)\equiv |d_0|e^{i\phi }$:
\begin{equation}
g=\sum_{m}\frac{2T_{m}^{2}}{(2-T_{m})^{2}-4(1-T_{m})(\epsilon _{\text{xc}%
}/|d_{0}|)^{2}}  \label{gs}
\end{equation}%
and $p=0$. The mixing conductance (\ref{gmix}) of $s$-wave magnetic
superconductors can also be expressed in terms of the scattering matrix of
the disordered region, by generalizing the formalism of Ref.~\cite%
{Beenakker95}. For that we need to concatenate the transfer matrix of the
normal disordered region with electron$\leftrightarrow $hole conversion at
the interface: $r_{eh(he)}^{\uparrow }=e^{i(\beta \pm \phi )}$ and $%
r_{eh(he)}^{\downarrow }=e^{-i(\beta \mp \phi )}$, where $\beta =\arccos
(\epsilon _{\text{xc}}/|d_{0}|)$. The situation is even more complicated
for the $\mathbf{k}$-dependent pair potential (which is always the case for
the odd-parity triplet pairing) and we do not pursue it here. The exception
is the case with no disordered region in our model, Fig.~\ref{sc}. The
conductance parameters for the singlet and triplet OSP in such ballistic systems are then given by 
\begin{eqnarray}
g &=&\frac{Ak_{F}^{2}}{2\pi}\,,\,\,p=0\,,\,\,\mathrm{and} \\
g^{\uparrow \downarrow } &=&A\int \frac{d^{2}\mathbf{k}_{\perp }}{(2\pi )^{2}%
}\left\{ 1\pm \exp \left[ -2i\arccos \frac{\epsilon _{\text{xc}}}{\Delta(\mathbf{k})}\right] \right\} \,,  \label{gOSP}
\end{eqnarray}%
where the $\pm$ sign corresponds to singlet (triplet) pairing and $\Delta=|d_0|$ or $|d|$, respectively. $\mathbf{k}$ is the
incident-electron wave vector at the Fermi level, $|\mathbf{k}|=k_{F}$, and $%
\mathbf{k}_{\perp }$ is its transversal projection in the lead with cross
section $A$. $r_{he}^{\uparrow }=e^{i(\beta -\phi )}$ and $%
r_{he}^{\downarrow }=-e^{-i(\beta +\phi )}$, where $\beta =\arccos (\epsilon
_{\text{xc}}/|d|)$ for the triplet OSP with $d=|d|e^{i\phi}$,
for a given $\mathbf{k}$.

It is instructive to discuss the mixing conductance $g^{\uparrow \downarrow }
$ (\ref{gmix}) values in some special cases of a clean interface with
matched band structures since it can be measured experimentally in FMR. For
an F/N interface, $r_{ee}^{\sigma }=r_{he}^{\sigma }=0$, while $g^{\uparrow
\downarrow }=M$ is large determined by the number of transverse wave-guide
channels $M$, assuming that the F layer is thicker than the ferromagnetic
coherence length $\hbar v_{F}/(\pi\epsilon_{\text{xc}})$. For a perfect
electron$\rightarrow $hole reflection off the singlet superconductor, we
find $r_{he}^{\uparrow }=-r_{he}^{\downarrow }$, $|r_{he}^{\sigma }|=1$, in
the limit when $\epsilon _{\text{xc}}\ll\Delta$, resulting in a
vanishing mixing conductance, as follows from Eq.~(\ref{gOSP}). This is easy
to understand since $r_{he}^{\uparrow }=-r_{he}^{\downarrow }$ means that
the transverse spin-$\uparrow $ electrons get reflected as the spin-$%
\downarrow $ holes which exactly cancel the incident transverse spin
current. In the analogous limit of the triplet OSP, $r_{he}^{\uparrow
}=r_{he}^{\downarrow }$, $|r_{he}^{\sigma }|=1$, doubling the F/N mixing
conductance. We thus find that $g^{\uparrow \downarrow }=M,\,0,\,2M$ for
F/N, singlet FS/N, and triplet OSP FS/N interfaces. Since $g^{\uparrow
\downarrow }$ is a direct measure of the ferromagnetic Gilbert-damping
enhancement \cite{Tserkovnyak:prl021}, the onset and the nature of the
superconducting pairing has nontrivial consequences in FMR experiments \cite%
{Fainstein:prb99}. It is worthwhile noting that $g^{\uparrow \downarrow
}\rightarrow 2M$ is a rather special limit for the mixing conductance in
multilayer magnetoelectronic circuit theories \cite%
{Bauer:prb03,Heinrich:prl03}. For example, in the case of a symmetric
FS/N/FS structure, it should result in a dynamic locking of the two
magnetizations which can be measured experimentally \cite{Heinrich:prl03}.

Finally, in triplet ESP with no disordered region, 
\begin{equation}
g=\frac{Ak_{F}^{2}}{4\pi}\,,\,\,p=\sigma \,,\,\,\mathrm{and}\,\,g^{\uparrow
\downarrow }=g
\end{equation}%
in the $A_{1}^{\sigma }$ phase, and 
\begin{eqnarray}
g &=&\frac{Ak_{F}^{2}}{2\pi}\,,\,\,p=0\,,\,\,\mathrm{and} \\
g^{\uparrow \downarrow } &=&A\int \frac{d^{2}\mathbf{k}_{\perp }}{(2\pi )^{2}%
}\left\{ 1+\exp \left( i[\phi ^{\uparrow }(\mathbf{k})-\phi ^{\downarrow }(%
\mathbf{k})]\right) \right\}   \label{A2mix}
\end{eqnarray}%
in the $A_{2}$ phase. $r_{he}^{\sigma }=ie^{-i\phi ^{\sigma }}$, for each
superconducting spin channel with $d^{\sigma }=|d^{\sigma
}|e^{i\phi ^{\sigma }}\neq 0$. Note that in deriving Eq.~(\ref{Is}), we made a
convention for the instantaneous coordinate system of $\mathbf{\hat{x}}%
\propto \mathbf{m}\times \partial _{t}\mathbf{m}$, $\mathbf{\hat{y}}\propto
-\partial _{t}\mathbf{m}$, and $\mathbf{\hat{z}}=\mathbf{m}$. [It is
necessary to specify the coordinate-system convention
in the case of the $A_{2}$ phase, because $d^{\uparrow (\downarrow
)}=\mp d_{x}+i d_{y}$ and, since $\boldsymbol{d}$
transforms as a vector, the relative phase of $d^{\uparrow (\downarrow
)}$ entering Eq.~(\ref{A2mix}) depends on the choice of the $x$ and $y$
axes.] It then follows that the second term in the curly brackets of
Eq.~(\ref{A2mix}) is modulated for a small-angle precession of $\mathbf{m}$,
unless $\boldsymbol{d}$ twists around $\mathbf{m}$
with its precession frequency
(i.e., following the instantaneous rotation axis). In the
former case, the theory therefore predicts an anisotropic Gilbert-damping
parameter. For the triplet ESP $A_{2}$ phase, the symmetry of the
superconducting order parameter may thus be seen in an additional anisotropic
FMR line width, as one changes the temperature across the F-to-FS transition.

In summary, we have studied the interplay between ferromagnetism and
superconductivity in adiabatic pumping by varying order parameters in FS
materials in contact with normal metals. We demonstrate that the symmetry of
the superconducting pair potential is reflected in the
conductance parameters which govern the pumped spin and charge flows. An
experimental quantity of a particular interest, which encodes information
about both ferromagnetic and superconducting correlations, is the mixing
conductance $g^{\uparrow \downarrow }$ which governs the Gilbert damping of
the magnetization dynamics \cite{Tserkovnyak:prl021}. Consequently,
signatures of the FS order parameter can be measured in thin film FS/N or
FS/N/FS FMR experiments.

We are grateful to G. E. W. Bauer, B. I. Halperin and A. Sudb\o for stimulating
discussions. This work was supported in part by the Research Council of Norway, NANOMAT Grants No. 158518/143 and 153458/431 and the Harvard Society of Fellows.


\begin{thebibliography}
\fi
\expandafter\ifx\csname bibnamefont\endcsname\relax

\fi
\expandafter\ifx\csname bibfnamefont\endcsname\relax

\fi
\expandafter\ifx\csname citenamefont\endcsname\relax

\fi
\expandafter\ifx\csname url\endcsname\relax

\fi
\expandafter\ifx\csname urlprefix\endcsname\relax

\fi
\providecommand{\bibinfo}[2]{#2} \providecommand{\eprint}[2][]{\url{#2}}

\bibitem[Tallon et~al.(1999)Tallon, Berhard, Bowden, Bilgerd, Soto, and
Pringle]{Tallon:ieee99} \bibinfo{author}{\bibfnamefont{J.}~%
\bibnamefont{Tallon} \textit{et al.}}, 
\bibinfo{journal}{IEEE Trans. Appl.
Supercond.} \textbf{\bibinfo{volume}{9}}, \bibinfo{pages}{1696} (%
\bibinfo{year}{1999}).

\bibitem[Saxena et~al.(2000)Saxena, Agarwal, Ahilan, Grosche, Haselwimmer,
Steiner, Pugh, Walker, Julian, Monthoux et~al.]{Saxena:nat00} %
\bibinfo{author}{\bibfnamefont{S.~S.} \bibnamefont{Saxena} \textit{et al.}}, %
\bibinfo{journal}{Nature} \textbf{\bibinfo{volume}{406}}, %
\bibinfo{pages}{587} (\bibinfo{year}{2000}).

\bibitem[Pfleiderer et~al.(2001)Pfleiderer, Uhlarz, Hayden, Vollmer, v.~L{\"o%
}hneysen, Bernhoeft, and Lonzarich]{Pfleiderer:nat01} \bibinfo{author}{%
\bibfnamefont{C.}~\bibnamefont{Pfleiderer} \textit{et al.}}, %
\bibinfo{journal}{Nature} \textbf{\bibinfo{volume}{412}}, \bibinfo{pages}{58}
(\bibinfo{year}{2001}).

\bibitem[Aoki et~al.(2001)Aoki, Huxley, Resouche, Brithwaite, Floquet,
Brison, Lhotel, and Paulsen]{Aoki:nat01} \bibinfo{author}{\bibfnamefont{D.}~%
\bibnamefont{Aoki} \textit{et al.}}, \bibinfo{journal}{Nature} \textbf{%
\bibinfo{volume}{413}}, \bibinfo{pages}{613} (\bibinfo{year}{2001}).

\bibitem[Huertas-Hernando et~al.(2002)Huertas-Hernando, Nazarov, and Belzig]%
{Huertas:prl02} \bibinfo{author}{\bibfnamefont{D.}~%
\bibnamefont{Huertas-Hernando}}, 
\bibinfo{author}{\bibfnamefont{Yu.~V.}
\bibnamefont{Nazarov}}, and \bibinfo{author}{\bibfnamefont{W.}~%
\bibnamefont{Belzig}}, \bibinfo{journal}{Phys. Rev. Lett.} \textbf{%
\bibinfo{volume}{88}}, \bibinfo{pages}{047003} (\bibinfo{year}{2002}).

\bibitem{GMROptics} S. Maekawa and T. Shinjo (Eds.), \textit{Applications of
Magnetic Nanostructures}, (Taylor and Francis, New York, U.S.A, 2002); D.D.
Awschalom, D. Loss, N. Samarth (Eds.), \textit{Semiconductor Spintronics and
Quantum Computing}, (Springer, Berlin, 2002).

\bibitem[Fainstein et~al.(2001)]{Fainstein:prb99} \bibinfo{author}{%
\bibfnamefont{A.}~\bibnamefont{Fainstein} \textit{et al.}}, %
\bibinfo{journal}{Phys. Rev. B} \textbf{\bibinfo{volume}{60}}, %
\bibinfo{pages}{R12597} (\bibinfo{year}{1999}).

\bibitem[Bruder(1990)]{Bruder:prb90} \bibinfo{author}{\bibfnamefont{C.}~%
\bibnamefont{Bruder}}, \bibinfo{journal}{Phys. Rev. B} \textbf{%
\bibinfo{volume}{41}}, \bibinfo{pages}{4017} (\bibinfo{year}{1990}).

\bibitem[Sigrist and Ueda(1991)]{Sigrist:rmp91} \bibinfo{author}{%
\bibfnamefont{M.}~\bibnamefont{Sigrist}} and \bibinfo{author}{%
\bibfnamefont{K.}~\bibnamefont{Ueda}}, \bibinfo{journal}{Rev. Mod. Phys.} 
\textbf{\bibinfo{volume}{63}}, \bibinfo{pages}{239} (\bibinfo{year}{1991}).

\bibitem[Powell et~al.(2003)Powell, Annett, and Gy{\"o}rffy]{Powell:jpa03} %
\bibinfo{author}{\bibfnamefont{B.~J.} \bibnamefont{Powell}}, %
\bibinfo{author}{\bibfnamefont{J.~F.} \bibnamefont{Annett}}, and 
\bibinfo{author}{\bibfnamefont{B.~L.}
  \bibnamefont{Gy{\"o}rffy}}, \bibinfo{journal}{J. Phys. A: Math. Gen.} 
\textbf{\bibinfo{volume}{36}}, \bibinfo{pages}{9289} (\bibinfo{year}{2003}).

\bibitem[Beenakker(1995)]{Beenakker95} \bibinfo{author}{%
\bibfnamefont{C.~W.~J.} \bibnamefont{Beenakker}}, in \emph{%
\bibinfo{booktitle}{Mesoscopic Quantum Physics}}, edited by %
\bibinfo{editor}{\bibfnamefont{E.}~\bibnamefont{Akkermans} \textit{et al.}} (%
\bibinfo{publisher}{Elsevier Science B. V.}, \bibinfo{year}{1995}), p. %
\bibinfo{pages}{279}.

\bibitem[Brouwer(1998)]{Brouwer:prb98} 
\bibinfo{author}{\bibfnamefont{P.~W.}
\bibnamefont{Brouwer}}, \bibinfo{journal}{Phys. Rev. B} \textbf{%
\bibinfo{volume}{58}}, \bibinfo{pages}{R10135} (\bibinfo{year}{1998}).

\bibitem[Brataas et~al.(2000)Brataas, Nazarov, and Bauer]{Brataas:prl00} %
\bibinfo{author}{\bibfnamefont{A.}~\bibnamefont{Brataas}}, %
\bibinfo{author}{\bibfnamefont{Yu.~V.} \bibnamefont{Nazarov}}, and 
\bibinfo{author}{\bibfnamefont{G.~E.~W.}
  \bibnamefont{Bauer}}, \bibinfo{journal}{Phys. Rev. Lett.} \textbf{%
\bibinfo{volume}{84}}, \bibinfo{pages}{2481} (\bibinfo{year}{2000}).

\bibitem[Tserkovnyak et~al.(2002)Tserkovnyak, Brataas, and Bauer]%
{Tserkovnyak:prl021} \bibinfo{author}{\bibfnamefont{Y.}~%
\bibnamefont{Tserkovnyak}}, \bibinfo{author}{\bibfnamefont{A.}~%
\bibnamefont{Brataas}}, and 
\bibinfo{author}{\bibfnamefont{G.~E.~W.}
\bibnamefont{Bauer}}, \bibinfo{journal}{Phys. Rev. Lett.} \textbf{%
\bibinfo{volume}{88}}, \bibinfo{pages}{117601} (\bibinfo{year}{2002}).

\bibitem[Blaauboer(2002)]{Blaauboer:prb02} \bibinfo{author}{%
\bibfnamefont{M.}~\bibnamefont{Blaauboer}}, \bibinfo{journal}{Phys. Rev. B} 
\textbf{\bibinfo{volume}{65}}, \bibinfo{pages}{235318} (\bibinfo{year}{2002}%
).

\bibitem[Bauer et~al.(2003)Bauer, Tserkovnyak, Huertas-Hernando, and Brataas]%
{Bauer:prb03} 
\bibinfo{author}{\bibfnamefont{G.~E.~W.} \bibnamefont{Bauer}
\textit{et al.}}, \bibinfo{journal}{Phys. Rev. B} \textbf{%
\bibinfo{volume}{67}}, \bibinfo{pages}{094421} (\bibinfo{year}{2003}).

\bibitem[Heinrich et~al.(2003)Heinrich, Tserkovnyak, Woltersdorf, Brataas,
Urban, and Bauer]{Heinrich:prl03} \bibinfo{author}{\bibfnamefont{B.}~%
\bibnamefont{Heinrich} \textit{et al.}}, \bibinfo{journal}{Phys. Rev. Lett.} 
\textbf{\bibinfo{volume}{90}}, \bibinfo{pages}{187601} (\bibinfo{year}{2003}%
).
\end{thebibliography}
\end{document}